\documentclass[reprint,amsmath,amssymb,aps, prb]{revtex4-2}
\usepackage{hyperref}
\usepackage[english]{babel} 
\usepackage{amssymb}
\usepackage{amsmath}
\usepackage{txfonts}
\usepackage{mathdots}
\usepackage[classicReIm]{kpfonts}
\usepackage{graphicx}
\usepackage[dvipsnames]{xcolor}
\def\arrvline{\hfil\kern\arraycolsep\vline\kern-\arraycolsep\hfilneg}
\usepackage{dcolumn}
\usepackage{bm}

\begin{document}

\title{Simultaneous measurements of nuclear spin heat capacity, temperature and relaxation in GaAs microstructures. }

\author{M. Vladimirova, S. Cronenberger, A.  Colombier, D. Scalbert}
\affiliation{Laboratoire Charles Coulomb, UMR 5221 CNRS/Universit\'{e} de Montpellier, France}
\author{V. M. Litvyak, K. V. Kavokin}
\affiliation{Spin Optics Laboratory, St.~Petersburg State University, Ulyanovskaya 1, St. Petersburg 198504, Russia}

\author{A. Lema\^itre}
\affiliation{ Universit\'{e} Paris-Saclay, CNRS, Centre de Nanosciences et de nanotechnologies, 91120, Palaiseau, France}

\begin{abstract}
 Heat capacity of the nuclear spin system (NSS) in GaAs-based microstructures has been shown to be much greater than expected from dipolar coupling between nuclei, thus limiting the efficiency of NSS cooling by adiabatic demagnetization.   It was suggested that quadrupole interaction induced by some small residual strain could provide this additional reservoir for the heat storage. We check and validate this hypothesis by combining nuclear spin relaxation measurements with adiabatic re-magnetization  and  nuclear magnetic resonance experiments, using  electron spin noise spectroscopy as a unique  tool for detection of nuclear magnetization. Our results confirm and quantify the role  of the quadrupole splitting in the heat storage within NSS and provide additional insight into fundamental, but  still actively debated relation between a mechanical strain and the resulting  electric field gradients in GaAs. 
\end{abstract}

\date{\today}
\maketitle

\section{Introduction} 
In n-doped semiconductors two spin ensembles, donor-bound electron spins and the  lattice nuclei
spin system (NSS) are coupled via hyperfine interaction \cite{OO}. 
Their magnetic moments,  interaction energies, their coupling to light, to crystal lattice vibrations, as well as to electric and magnetic fields differ dramatically.
The resulting physics, that is best studied in n-GaAs, is  quite complex.
However, if sufficient NSS stability is reached, it may potentially offer novel applications for quantum information technologies \cite{Foletti2009,Bluhm2010,Chekhovich2013,Urbaszek2013,Stockill2016,Gangloff2019,Denning2019,Chekhovich2020,Gangloff2021}.

%
%
%
%
\begin{figure}
	\includegraphics[width=3.2in]{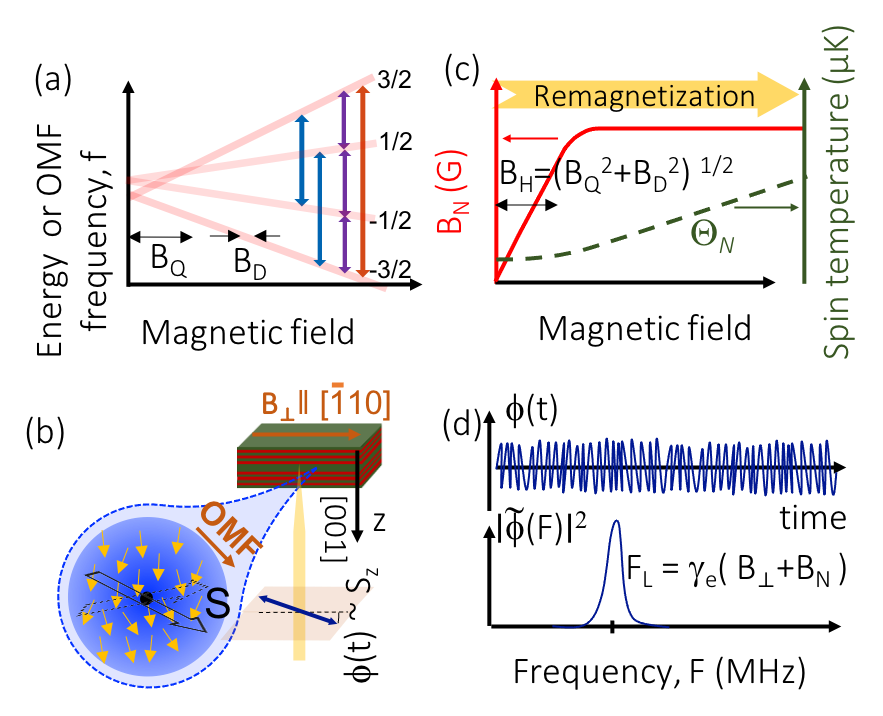}
	\caption{  (a)  Schematic  representation of  quadrupole-split nuclear spin levels.
	Arrows indicate NMR transitions, (b) Sketch of the donor-bound electron spin interacting with the underlying nuclear spins.   (c) Overhauser field $B_N$ and the corresponding nuclear spin temperature  $\Theta_N$ as functions of the magnetic field according to spin temperature theory. $B_H$ is a  field that characterizes NSS heat capacity.   (d) Fluctuations of the electron spin detected via Kerr rotation in time domain within $0.5$~GHz frequency band and the corresponding Fourier power spectrum.
	}
		\label{fig:fig1}
\end{figure}

In this context, precise knowledge and control of nuclear quadrupole effects in  n-GaAs and GaAs-based heterostructures has attracted substantial attention.
Indeed, in semiconductors with cubic symmetry containing isotopes with spin $I>1/2$  quadrupole interaction induced by  strain splits NSS energy states ({\it cf} Fig.~\ref{fig:fig1}~(a)) and thus strongly affects NSS thermodynamics \cite{Abragam,Dzhioev2007,Maletinsky09,Chekhovich2012}. 
%
{
The effects of strain are particularly important in heterostructures, such as quantum wells \cite{Kotur2021} and quantum dots (QDs), and were extensively studied in both single QDs \cite{Chekhovich2013,Chekhovich2018} and QDs ensembles \cite{Flisinski2010}.}
More recently quadrupole effects in bulk n-GaAs have been evidenced, suggesting the importance of the precise control of both strain and electric field when NSS needs to be efficiently cooled down \cite{Litvyak2021}.
Moreover, 
the parameters of different isotopes determining the relation between the strain tensor elements and the resulting gradients of electric fields  remain  controversial \cite{Sundfors1974,Chekhovich2018,Griffiths2019,Litvyak2021}.
This is partly due to the lack of non-destructive and non-perturbative experimental techniques, capable of probing nuclear magnetic resonances (NMR) at low and zero magnetic field, where quadrupole splittings between spin levels dominate over Zeeman ones. %
%

{Indeed, traditional optically detected NMR experiments addressing QD ensembles are limited to the field range given by the Hanle curve, and suffer from huge broadening due to  inhomogeneity of the electric field gradient experienced by the nuclei \cite{Flisinski2010}. 
On the other hand, since pioneering experiments of Gammon et al resolving PL fine structure in individual QDs \cite{Gammon1996},   remarkable progress has been achieved in the NSS control in such structures \cite{Chekhovich2013}. However, in order to reach measurable splitting a relatively high magnetic field, where Zeeman effect dominates over quadrupole one, should be applied.
 }

Spin noise spectroscopy (SNS) could be promising for this purpose. It has been shown that electron spin noise can non-destructively relay properties of nuclear environment through the statistics of electron spin interaction with the total nuclear polarization \cite{Glazov2015,Ryzhov2015,Berski2015,CH11Spin2017,Glazov2018}. 
SNS  is based on  the non-perturbative detection of the spin polarization noise via fluctuations of the Kerr or Faraday rotation of the off-resonant probe beam polarization Fig.~\ref{fig:fig1}~(b)\cite{Zapasskii,Romer2007}.

{
It is important to stress that although we aim at probing the NSS, rather than electron spin, what is measured in practice is the electron spin noise spectrum in the megahertz frequency range.  
This approach differs from the one adopted by Berski et al \cite{Berski2015}, who measured in n-GaAs directly nuclear spin noise associated with different isotopes in the kilohertz frequency range. 
Such direct nuclear spin measurements are quite difficult, since NSS does not couple directly with light, but they are particularly useful to address the nuclear spin dynamics at low magnetic fields hardly accessible otherwise.
}

In electron SNS, the fluctuation spectrum  exhibits a pronounced peak at the Larmor frequency  corresponding to the total magnetic field experienced by electrons. It is given the sum of the external field $B_\perp$ and Overhauzer field $B_N$, an effective nuclear field resulting from the hyperfine interaction, see Fig.~\ref{fig:fig1}~(d).
Therefore any  variation of the NSS polarization, either due to relaxation or when external magnetic field is reduced, results in a change of the Larmor frequency and thus in a shift of the electron spin noise peak. This shift is proportional to  the Overhauser field, but is not isotope-selective. We have demonstrated that evolution of nuclear spin polarization during optical pumping, relaxation and  re-magnetization through  zero field can be monitored via this shift with sub-second resolution \cite{Ryzhov2015,CH11Spin2017,Vladimirova2017,Vladimirova2018}.

These experiments performed on in n-GaAs layers revealed that   the relaxation rates of optically cooled NSS under magnetic fields below $10$~G are dramatically enhanced.
We also measured  huge heat capacity characterized by so-called local field, also of order of  $\approx 10$~G.
This value is  much higher than expected from magnetic dipole-dipole interaction $B_D\approx 1.5$~G \cite{Paget1977}.
This result is illustrated in Fig.~\ref{fig:fig1}~(b).
It shows how the Overhauser field (solid line) and the nuclear spin temperature (dashed line) in a cold NSS vary during re-magnetization from zero to high magnetic field. 
%
{
The dependence of the Overhauser field on the magnetic field $B$ is related to the nuclear spin temperature $\Theta_N$ \cite{OO}:
\begin{equation}
 B_N=\frac{ h B b_N    I(I+1)\bar{\gamma}_N } {3k_B \Theta_N}.
\label{eqSpinT0}
\end{equation}
%
Here $\bar{\gamma}_N=\sum_i { A_i \gamma_{Ni}}$ is the average gyromagnetic ratio, $A_i$ ($\gamma_{Ni}$) is the abundance (gyromagnetic ratio) of $i$-th isotope (see Table~\ref{tab:table1}), \textit{h} and $k_B$ are Planck and Boltzmann constants, respectively, $b_N$ is the Overhauser field at saturation of the nuclear magnetization, in GaAs $b_N=5.3$~T.
}
The magnetic field  below which $\Theta_N$ remains constant reveals the  heat capacity of the NSS. In the following it will be referred to as the heat capacity field, $B_H$ \footnote{This field was sometimes called local field despite the quadrupole contribution \cite{Wolff}, but for the sake of clarity we prefer using the term heat capacity field}. The value $B_H>>B_D$ as shown in Fig.~\ref{fig:fig1}~(c) is a fingerprint of the enhanced NSS heat capacity as compared to the traditional spin temperature theory \cite{OO,Vladimirova2018}. 

Because in GaAs the spin of all isotopes $I=3/2$, it was suggested that such an enhanced NSS heat capacity could be explained by the quadrupole splitting of nuclear spin energy levels induced by some small residual strain, see  Fig.~\ref{fig:fig1}~(b). The corresponding  effective field $B_Q$ is the one where Zeeman and quadrupole energies become comparable.
In re-magnetization experiment illustrated in Fig.~\ref{fig:fig1}~(a) one would expect 
\begin{equation}
B_H=\sqrt{B_D^2+B_Q^2}. 
\label{eq:BH}
\end{equation}

\begin{figure}
	\includegraphics[width=3.4in]{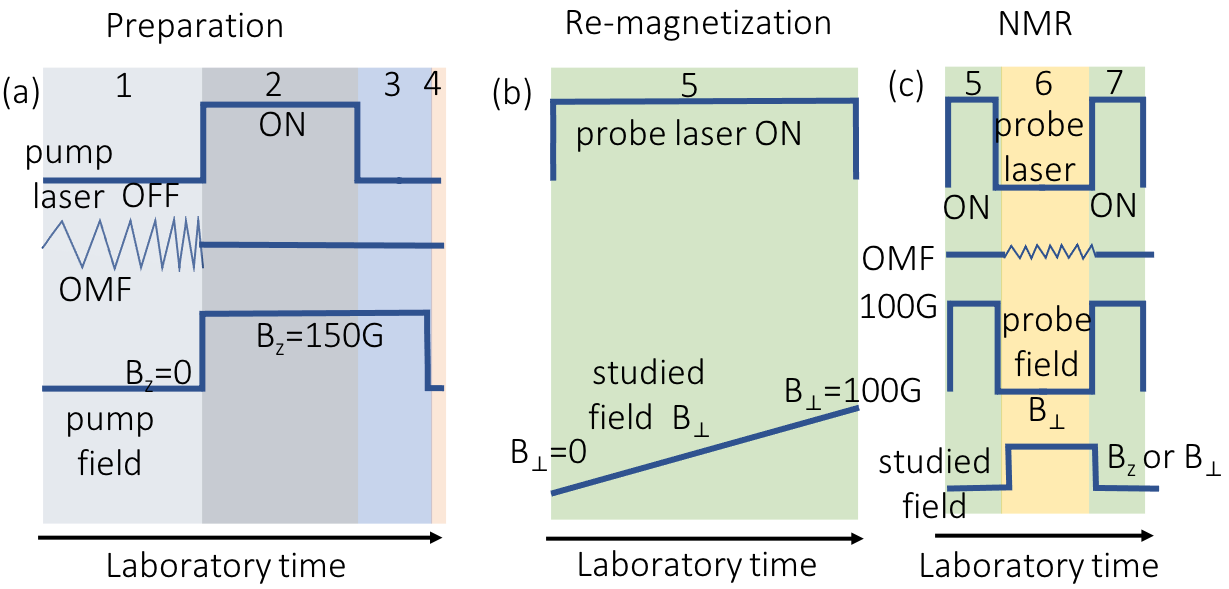}
	\caption{ 
{Time-sequence used for NSS preparation (a) and re-magnetization (b) and NMR (c) experiments.
}
}
	\label{fig:Protocol1}
\end{figure}
\begin{figure}
	\includegraphics[width=3.4in]{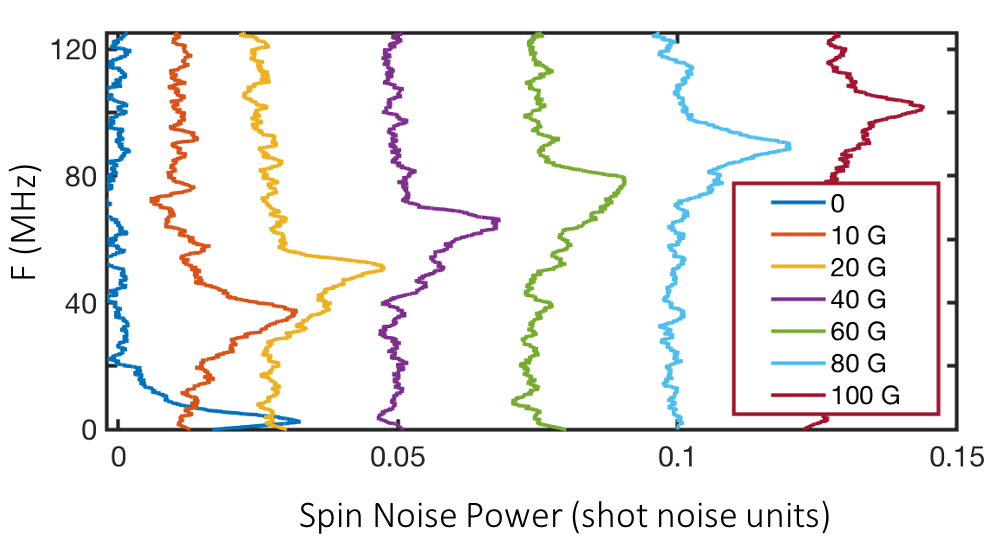}
	\caption{ 
{ SN spectra measured at different values of the magnetic field during adiabatic re-magnetization. Arrows point the corresponding Larmor frequencies $F_L$.}
}
	\label{fig:ProtocolBird}
\end{figure}
\begin{figure}
	\includegraphics[width=3.4in]{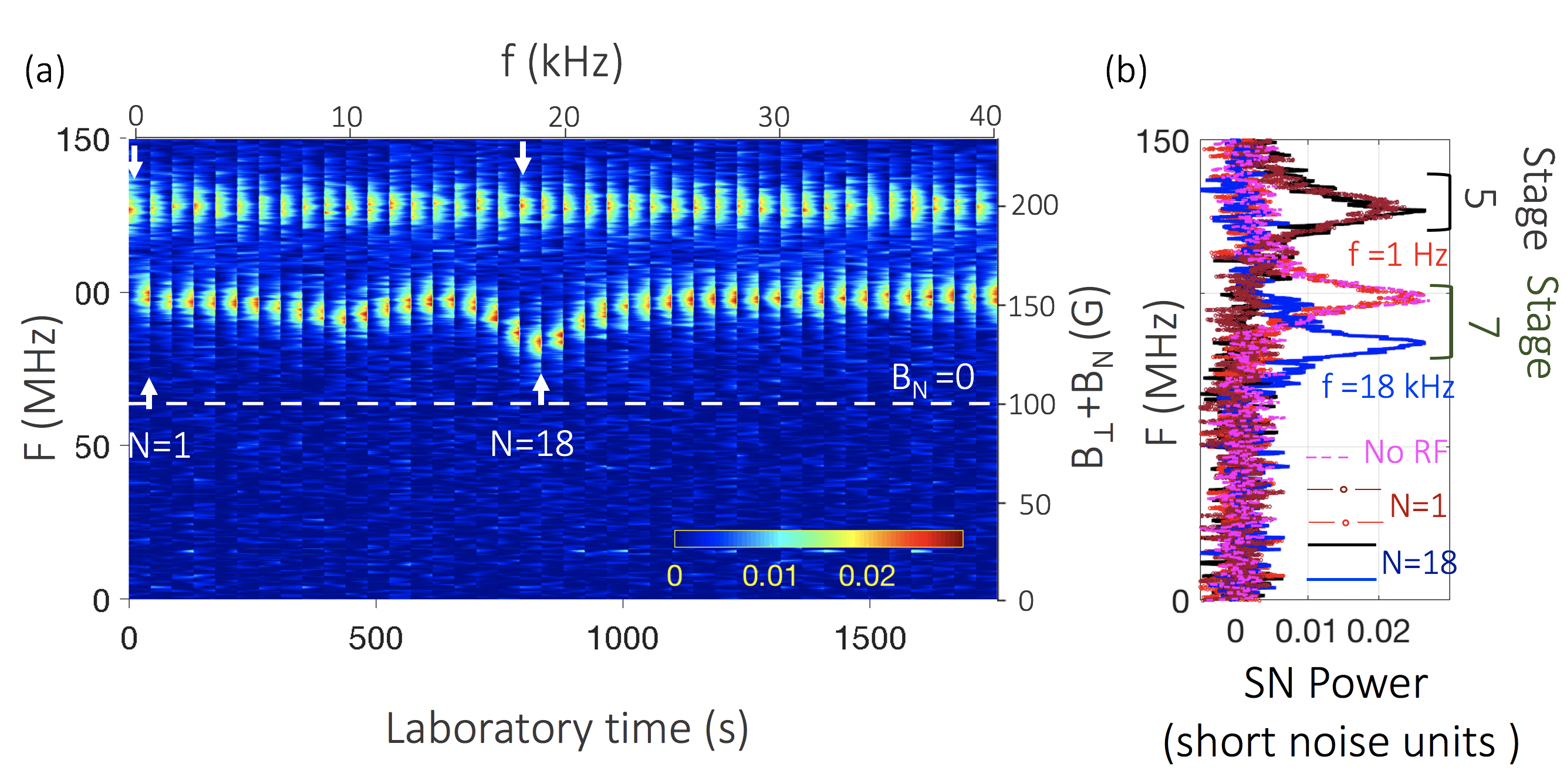}
	\caption{
{(a) $40$ pairs of color-encoded SN spectra required for NSS absorption measurement at $B=0$. Each pair is recorded within steps 5 and 7 of the NMR experiment that includes both preparation and detection. The measurements are realised in a single run one after another for different OMF frequencies $f$. Bottom scale indicates the laboratory time, and the top scale the corresponding OMF frequency. Such set of measurements allows us to reconstruct the NMR spectrum at $B=0$. Intensities are given in the units of spin to shot noise ratio.
	(b) Pairs $N=1$ ($f=1$~Hz) and $N=18$ ($f=18$~kHz) of the spin noise spectra shown in (a). At $f=1$~Hz NSS absorbs energy very weakly, so that the corresponding stage 7 spectrum is identical to the one measured in the absence of the OMF (dashed magenta line) and its peak position is essentially governed by NSS relaxation. The peak of the spectrum corresponding to $N=18$ is shifted to much lower frequency, this shift is  governed by NSS absorption.
	}
	}
	\label{fig:ProtocolRMN}
\end{figure}
However, in the absence of  NMR experiments that could provide direct access to quadrupole energies these ideas remained unconfirmed.

In this work we  implement a new technique, NMR detected by electron SNS, that allows us to confirm this hypothesis.
NSS absorption spectra in the radio-frequency (RF) range  are measured as a function of magnetic field from zero to $100$~G.
In order to reach the detection sensitivity required for such experiments (sub-second integration times) we use n-GaAs layer embedded in a microcavity \cite{Giri2013} and adopt the homodyne detection scheme for the for spin noise measurements \cite{Cronenberger2016,Sterin2018,Petrov2018}. 
The analysis of the NMR spectra allows us  to (i) evaluate quantitatively quadrupole energies and confirm the impact of the quadrupole interaction on the heat capacity of the NSS. The quadrupole energies extracted from the NMR spectra appear to be consistent with heat capacity fields measured in the NSS re-magnetization experiments and with the spin temperature theory; (ii) to provide additional insight into relative values of so-called gradient-elastic tensor $S_{ijkl}$ for Ga and As isotopes.
This  fundamental parameter determines the relation between the strain tensor $\epsilon_{kl}$ and the resulting gradients of electric fields at each atomic site $V_{ij}=\sum_{k,l} S_{ijkl}\epsilon_{kl}$.  Its values are isotope-dependent and are still actively debated in the literature \cite{Sundfors1974,Chekhovich2018,Griffiths2019,Litvyak2021}.
\begin{figure*}
	\includegraphics[width=6.0in]{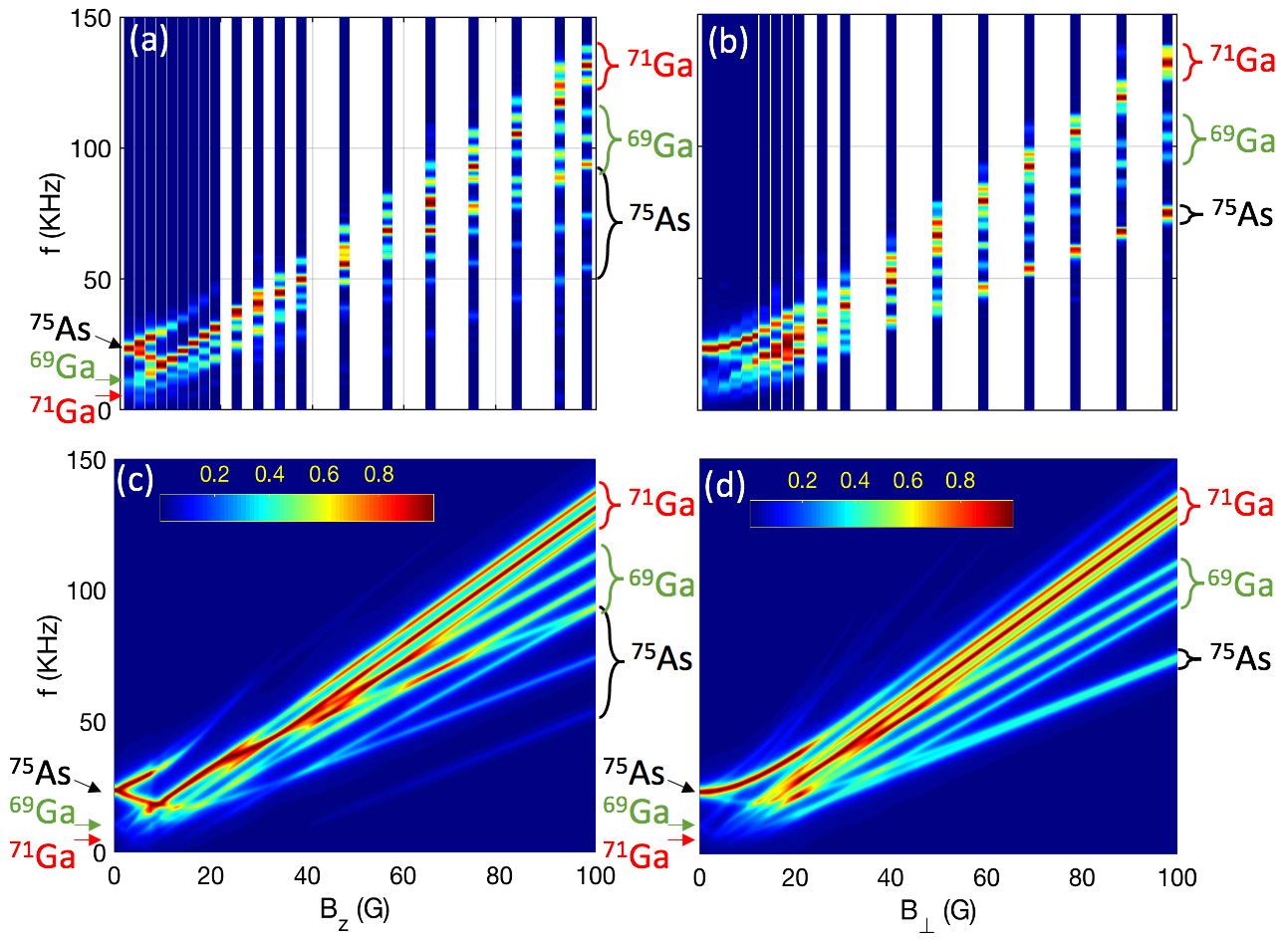}
	\caption{ Color-encoded NSS absorption spectra  measured (a, b) and calculated (c, d) at different values of magnetic field  parallel to either $[\bar{1}10]$ (b) or $[001]$ (a) crystallographic axis. All the spectra are normalized to unity.}
	\label{fig:ColorMap}
\end{figure*}
\section{Sample} 
We study the  microcavity sample cut out from the same wafer as in Ref. \onlinecite{Vladimirova2018}: a Si-doped $3\lambda/2$ GaAs  layer (cavity mode at $\approx 830$~nm in the studied piece of the sample ) with donor concentration $n_D\approx 2 \times 10^{15}$~cm$^{-3}$ sandwiched between two Bragg mirrors, in order to enhance the sensitivity of the SNS (quality factor $\approx 2\times10^4$). The front (back) mirrors are distributed Bragg reflectors composed of $25$ ($30$) pairs of AlAs/Al$_{0.1}$Ga$_{0.9}$As layers. The sample is grown on a $400$-$\mu$m-thick $[001]$-oriented GaAs substrate and placed in a cold finger cryostat at T=$6$~K.
A $30$-spires copper coil is  placed on top of the sample to create an oscillating magnetic field (OMF). It is directed at $45^o$ out of the sample plane \footnote {The intended direction of the OMF was along $z$-axis. Its in-plane component is presumably due to eddy currents in the copper-based sample holder.}.
\section{NMR and re-magnetization  detected by the electron spin noise} 
\label{sec:protocol}
The principle of  NMR and NSS re-magnetization detected by electron spin noise are sketched in Fig.~\ref{fig:Protocol1}. 
Prior to the measurement, any preexisting polarization in the NSS is erased by OMF $B_{RF}=0.3$~G during $20$~s, its frequency sweeping between $0$ and $50$~kHz (stage 1).  At second stage the NSS is polarized by optical pumping in the presence of a longitudinal magnetic field $B_z=150$~G during typically $30$s (circularly polarized laser diode emitting at $770$~nm). This is followed by $30$~s of dark time under the same magnetic field to allow for fast relaxation of the nuclear spins under the donors orbits while keeping bulk nuclei polarized (stage 3). { After that the NSS is  adiabatically demagnetized down to zero magnetic field (stage 4). This last stage is important to include in the protocol, despite some additional NSS warm-up that it causes. Indeed} it ensures NSS thermalization and a transfer of the optically pumped energy stored in Zeeman reservoir towards internal degrees of freedoms, that is dipole-dipole reservoir. {This additional nuclear spin relaxation during stage 4 does not play any role in the experiment, provided that NSS is strongly polarized (which is the case). What matters is to start the detection of the the NSS  demagnetisation or absorption each time at the same spin temperature.}

The four stages preparation of the NSS is identical for both NMR and re-magnetization experiments,Fig.~\ref{fig:Protocol1}~(a). 

The re-magnetization experiments are similar to those reported in Ref. \onlinecite{Vladimirova2018}. After preparation stages $1-4$,  a set of electron spin noise spectra under in-plane magnetic field that increases from $B=0$ to $B=100$~G is measured every $2$~G, see Fig.\ref{fig:Protocol1}~(b).  The measured frequency $F_L$ of the electron spin noise spectral peak at a given field $B_\perp$ is directly related to the Overhauser field $B_N$, 
\begin{equation}
F_L=F_B+F_N=\gamma_e(B_\perp+B_N),
\label{eq:freq}
\end{equation}
where $\gamma_e=0.62$~MHz/G is the electron gyromagnetic ratio in GaAs. Representative spectra measured at several values of magnetic field are shown in Fig.\ref{fig:ProtocolBird}. The corresponding Larmor frequencies  are pointed by arrows.

Measurement of each point of the NMR spectrum at a given field $B$ (either $B_\perp$ or $B_z$) corresponding to a given RF frequency $f$ consists in three steps illustrated in Fig.~\ref{fig:Protocol1}~(c): 
{detection of electron spin noise and its peak frequency $F_L$ in the sample under magnetic field $B_\perp=100$~G in order to evaluate the Overhauser field prior to absorption (stage 5), application during $3$~s of the OMF at a given frequency $f$ in the RF range  and at a given magnetic field $B$ at which the RF absorption is studied (stage 6), and again detection of electron spin noise and its frequency $F_L$ under magnetic field $B_\perp=100$~G (stage 7). This allows us to deduce from Eq.~\ref{eq:freq} to which extent the Overhauser field  has decreased during stage 6.
}

{Typical spin noise spectra measured at stages 5 and 7 are shown in Fig.\ref{fig:ProtocolRMN}~(b). The corresponding field under study is $B=0$. The OMF frequencies applied during stage 6 are $f=1$~Hz where NSS absorbs energy very weakly, and $f=18$~kHz where NSS absorption is strong. The spectrum measured during stage 7 in the absence of OMF is shown by dashed magenta line for comparison.
}
 The shift between the electron spin noise spectral peaks measured during stages 5 and 7 contains two contributions. The first is RF-independent, it is due to some weak, but unavoidable NSS relaxation during measurement. It can be evaluated by performing identical measurements in the absence of the  OMF during stage 6, 
{see dashed magenta line in Fig.\ref{fig:ProtocolRMN}~(b)}. The second contribution is proportional to the NSS absorption at a given OMF frequency $f$. It is negligibly small at $f=1$~Hz, but when  $f$ matches the energy difference between spin states of one of the isotopes (as for $f=18$~kHz, where $f$ matches $^{75}$As resonance) the radiation is efficiently absorbed 
{and the spectrum shifts more than in the absence of the OMF field}. 
 
The entire absorption spectrum (or NMR spectrum) can be constructed by repeating the protocol (stages 1-7) at different values of $f$. The details of this procedure are given in Appendix \ref{sec:appendixA}.
An example of the raw data obtained in a NMR experiment addressing NSS absorption at $B=0$ is shown in Fig.~\ref{fig:ProtocolRMN}~(a). Color-encoded set of 40 pairs of electron spin noise spectra measured in the units of the ratio between spin and shot noise are shown as a function of laboratory time (bottom scale). The corresponding OMF frequency $f$ is indicated on the top. For each value of $f$ two spectra are systematically measured, one before application of the OMF (in this case the electron spin noise peak frequency $F_L$ does not depend on the OMF frequency $f$), and one after. Left scale shows spin noise frequency and right scale the corresponding total magnetic field experienced by the electrons, $B_\perp+B_N=F/\gamma_e$, where $B_\perp=100$~G. One can see that $B_N$ is significantly smaller after application of the OMF at $f \approx 8$~kHz and at $f\approx 20$~kHz. These frequencies  characterize nuclear spin splittings at $B=0$, {\it cf} Fig.~\ref{fig:fig1}~(a). We anticipate, that the higher frequency is related to quadrupole-split  states of $^{75}$As which has the highest quadrupole moment, and the lower one to closely lying transitions of $^{71}$Ga and $^{69}$Ga isotopes.

 It is important to note that in most of SNS experiments the magnetic field is modulated to eliminate spin-independent signals \cite{Hubner2014}. Such approach is not suitable for studies of the NSS, because nuclear spins get depolarized by non-adiabatically varying magnetic fields.  To avoid this effect and also to increase the sensitivity, in this work we use a homodyne detection scheme \cite{Cronenberger2016}. This  allows us to use the probe power as low as $150$~$\mu$W focused on $30$~$\mu$m-diameter spot.  We detect the spectra in two mutually orthogonal linear polarizations, that is either parallel (which contains useful information about electron spin noise) or orthogonal to the local oscillator polarization. The spin noise spectrum is obtained by taking  the difference between these spectra, in order to get rid of the background noise.

\section{NMR spectra and their interpretation}
A set of color-encoded NMR spectra measured under in-plane static magnetic field $B_{\perp} \parallel [110]$ and at $B_z \parallel [001]$, is shown in Fig.~\ref{fig:ColorMap}~(a-b).
In order to understand these spectra in terms of the OMF-induced transitions between spin states of the three GaAs isotopes, we must calculate the energy spectrum of the NSS. It is determined by both quadrupole interaction and Zeeman effects: $\hat{H}^i=\hat{H}_Q^i+\hat{H}_{Z}^i$,
were $\hat{H}_Q^i$ ($\hat{H}_Z^i$) describes quadrupole (Zeeman) interaction \footnote{The dipole-dipole interaction which is known to broaden the NMR transitions, but does not lead to any shifts of their energies, is not included here\cite{Abragam}}.  The energy levels of each isotope are given by the eigenvalues of the Hamiltonian ${\hat{H}}^i$, and the NMR transitions frequencies by their differences $f_{kl}^i=(E_k^i-E_l^i)/h$. The OMF induces spin transitions between a pair of states  $k$ and $l$ of $i$-th isotope, if the OMF frequency matches the energy difference between  their energy levels $E_k$ and $E_l$. The probability of corresponding transition is given by  $P_{kl}^i \propto M^2_{kl,i}$, where $M_{kl}=\langle \Psi _k\left|H_{OMF}\right|\Psi_l\rangle$ is the matrix element of the Hamiltonian describing Zeeman interaction of the nuclear spins with the OMF.

The   quadrupole Hamiltonian of the $i$-th isotope in the presence of the  in-plane biaxial strain can be written as \cite{Abragam}:
\begin{equation} \label{eq:HQi} 
\begin{split}
\hat{H}_{Q}^i=&\frac{E_{QZ}^i}{2}
 \left[\hat{I}_z^2 -\frac{I(I+1)}{3}\right] +   \\
 & \frac{E_{QR}^i}{2\sqrt{3}}
 \left[
 \left(\hat{I}_x^2 - \hat{I}_y^2\right))\right]+ \frac{E_{QI}^i}{2\sqrt{3}}\left[ \left(\hat{I}_x\hat{I}_y + \hat{I}_y\hat{I}_x\right)\right],
\end{split}
\end{equation} 
where $E_{QR}^i=E_{Q\perp}^i \mathrm{cos}2\zeta^i$ and $E_{QI}^i=E_{Q\perp}^i  \mathrm{sin} 2\zeta^i$.
It is  determined by three parameters: 
$E_{QZ}^i$ - the quadrupole energy along growth axis, ${E_{Q\perp}^i}$ - the in-plane quadrupole energy, and $\zeta^i$ - the angle between the principal axis of the electric field gradient tensor $V_{ij}$ and the $[100]$ crystal axis.  
\begin{figure}
	\includegraphics[width=3.4in]{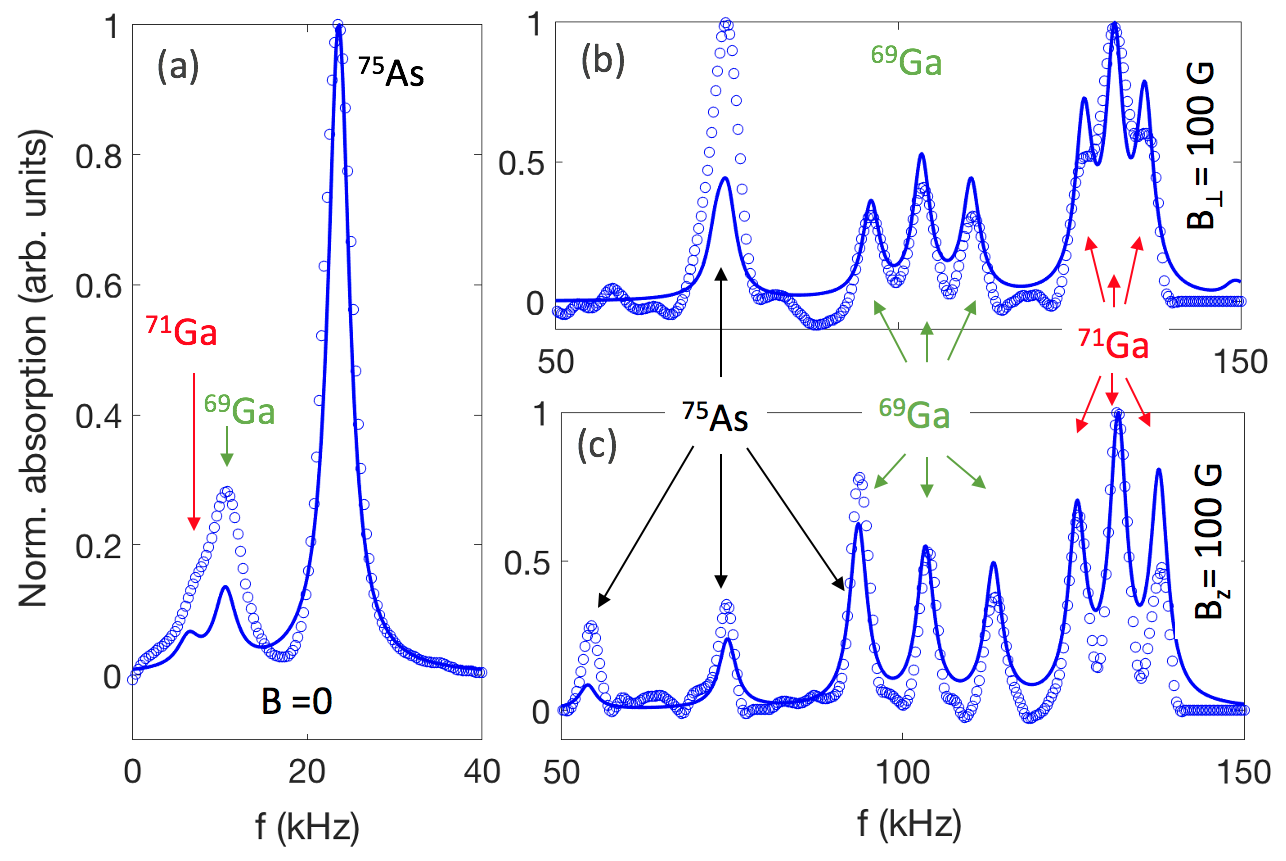}
	\caption{ 
{NSS absorption spectra  measured (symbols) and calculated (lines) at $B=0$ (a), at $B_z=100$~G, that is parallel to $[001]$ crystal axis (b) and at  $B_\perp=100$~G, parallel to $[\bar{1}10]$. All the spectra are normalized to unity.
}}
	\label{fig:RMNspectra}
\end{figure}

The ensemble of the absorption spectra shown in Fig.~\ref{fig:ColorMap}~(a-b) can be understood  by fitting this simple model to the observed transition frequencies. This yields  $3$  parameters per isotope. They are  summarized in Table~\ref{tab:table1} together with other relevant parameters (abundances, quadrupole moments and gyromagnetic ratios) taken from the literature. Assuming that all transitions have Lorenzian shape broadened by dipole-dipole interaction ($\Delta f_D=1.5$~kHz),  we calculate the resulting absorption spectra at different values and two orientations of the magnetic field as shown in Fig.~\ref{fig:ColorMap}~(c-d). 

One can see that most of salient features of the measured spectra are faithfully reproduced. 
%
{ Indeed, at $B=0$, see also measured and calculated spectra presented separately in Fig.~\ref{fig:RMNspectra}~(a), the $^{75}$As absorption (higher frequency line) dominates the spectrum. 
}
 At small fields parallel to the growth axis $B_z$ the  $^{75}$As absorption line splits into  two lines, that shift  linearly  with $B_z$. By contrast, under in-plane field $B_\perp$ the $^{75}$As absorption line shifts quadratically. 

At strong field $B=100$~G we identify for each isotope a triplet of lines, including a central line and a pair of satellites.
%
{
Measured and calculated spectra at $B=100$~G oriented either along  $[\bar{1}10]$ or  $[001]$ crystallographic axis are shown in Fig.~\ref{fig:RMNspectra}~(b) and (c), respectively.
}
The strongest absorption is observed for $^{71}$Ga. In the presence of the longitudinal field $B_z$ the splitting 
between central line and its satellites is the smallest for $^{71}$Ga and the highest for $^{75}$As. By contrast, under an in-plane field $B_\perp$ this splitting vanishes for $^{75}$As.
This later fact results from the sign difference in the in- ( $E_{QI}$) and out-of-plane ($E_{QZ}$ ) quadrupole parameters for Ga and As. We will show below that this implies a particular sign relation between different elements of the gradient-elastic tensor elements  $S_{11}$,  $S_{44}$ for Ga and As.
%

{
Two more observations could be drawn from the comparison between the model and measured spectra.  %
First, the generic Lorenzian broadening $\Delta f_D=1.5$~kHz common for all transition fits the data satisfactorily. The fact that this linewidth is identical for all transitions and comparable with the one expected from dipole-dipole broadening (see also Fig.~\ref{fig:fig1}~(a))  suggests that NSS is efficiently thermalized at $B \approx B_D$.  
Second, the relative intensities of Ga and As absorption are not faithfully reproduced by the model:  at low magnetic field it underestimates Ga absorption as compared to As, and at high field the situation is inverted. We have no solid physical explanation for this discrepancy, which was also reported in Ref.\onlinecite{Berski2015},  { but not in Ref.\onlinecite{Litvyak2021}.  
Nevertheless it could simply result from ill-defined direction of the OMF field.}
}

\section{Re-magnetization of the NSS in the presence of the quadrupole effects}
The above measurements show  that quadrupole interaction is much stronger in this sample than dipole-dipole interaction. 
In this section we compare quantitatively the quadrupole splittings and effective field $B_Q$ that they should induce with the actual heat capacity field $B_H$ characterizing NSS re-magnetization.
 
The spin temperature theory states that in a NSS isolated from the lattice  an equilibrium state characterized by the temperature $\Theta_N$ will be  established within a characteristic time $T_2\approx h/\gamma_NB_D$ \cite{AbragamProctor}.
On time scales  larger than $T_2$ and shorter than $T_1$, and provided that  $T_2 \ll T_1$, where $T_1$ is  spin-lattice relaxation time, the NSS can be considered as isolated from the lattice.
If NSS is  prepared at temperature $\Theta_{Ni}$   under magnetic field $B_i$ and subjected to a slowly varying magnetic field, such that $dB/dt<B_D/T_2$, then $\Theta_N$ changes obeying universal expression  \cite{AbragamProctor,JETP1982,Spin2017,Vladimirova2018}:
\begin{equation}
\frac{\Theta_N}{(B^2+B_H^2)^{1/2}}=\frac{\Theta_{Ni}}{B_i}.
\label{eqSpinT2}
\end{equation}
where $\Theta_N$ is related to $B_N$ via Eq.~\ref{eqSpinT0}.
%

\begin{figure}
	\includegraphics[width=3.2in]{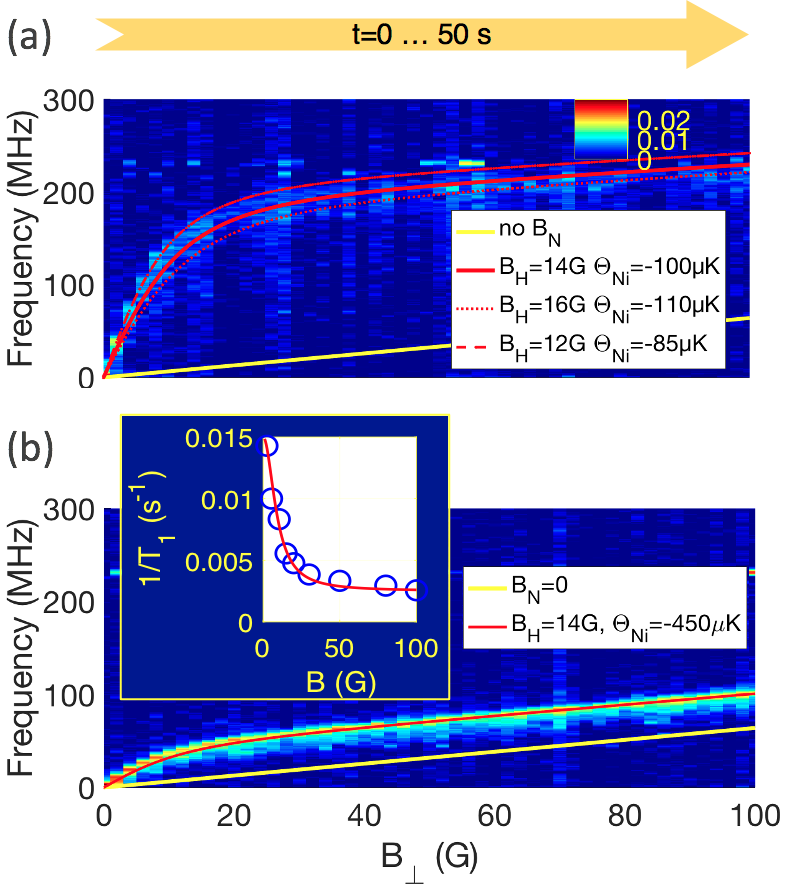}
	\caption{ Color-encoded SN spectra (in signal to shot noise ratio units) during the adiabatic demagnetization procedure preceded by  preparation of nuclei including either $4$~minutes (a) or  $30$~s optical pumping (b).   Red lines are fit to spin temperature model assuming $B_H=14\pm2$~G and taking into account spin relaxation. Yellow line shows the external field contribution to the spin noise frequency, $F_N$. Inset shows the inverse nuclear spin relaxation time as a function of the magnetic field (symbols) measured under the pumping conditions of (b). Solid line is a lorentzian fit to the data used to calculate red lines.  }
	\label{fig:demag}
\end{figure}

 \begin{table*}
\begin{tabular}{|c | c | c| c |c| }
\hline
 {} & ${}^{69}$Ga  &   ${}^{71}$Ga &   ${}^{75}$As  &   $^{75}$As/$^{ }$Ga  \\ 
\hline 
Q  (${10}^{-30}$ m${}^{2}$) \cite{doi:10.1080/00268970802018367} & 17.1  & 10.7 &  31.4 & {}  \\ \hline 
{Abundance}& {$0.3$}& {$0.2$}& {$0.5$ }& {} \\
\hline 
$\gamma_N $ (kHz/G) \cite{doi:10.1080/00268970802018367} & 0.64  &  0.82  &  0.45&{}  \\
\hline 
$C_{11}$ (${10}^{-10}$ ${\mathrm{N}}/{{\mathrm{m}}^{\mathrm{2}}})$ \cite{SS73} &   {12} &    12 & {12}&  \\ 
\hline 
$C_{12}$ (${10}^{-10}$ ${\mathrm{N}}/{{\mathrm{m}}^{\mathrm{2}}})$  \cite{SS73}&{5.4}   &   5.4 & {5.4}& \\ 
\hline 
$C_{44}$ (${10}^{-10}$ ${\mathrm{N}}/{{\mathrm{m}}^{\mathrm{2}}})$  \cite{SS73} &  {6.2} &    6.2& {6.2}& \\ 
\hline
$E_{QZ}$/h (kHz) & -10$\pm$0.5 & -6$\pm$0.5 &   20$\pm$0.5& {}  \\ 
\hline 
$E_{\perp}$/h (kHz)& -3.5$\pm$0.5 &   -2.2$\pm$0.5 &   12.5$\pm$0.5 & {} \\ 
\hline 
$\zeta$ (degrees with $[100]$ axis) & -68$\pm$5   &    -68$\pm$5  &   -33$\pm$5 & {}  \\
\hline 
$S_{11}$ ($ {10}^{21}  {\mathrm{V}}/{{\mathrm{m}}^{\mathrm{2}}}$ )  &  {-31.7}\cite{Sundfors1974}, -22 \cite{Chekhovich2018}   &  {-31.7}\cite{Sundfors1974}, -22 \cite{Chekhovich2018}   &   { 34}\cite{Sundfors1974}, 24.2 \cite{Chekhovich2018}     & {-1.07} \cite{Sundfors1974}, -1.1 \cite{Chekhovich2018,Litvyak2021}  \\ 
\hline 
$S_{11}$ ($^{75}$As) / $S_{11}$($^{ }$ Ga) {\bf[this work] }
&{} &  {}   &  {}    & {\bf -1.1$\pm$0.1}  \\ 
\hline 
$S_{44}$  ($ {10}^{21}  {\mathrm{V}}/{{\mathrm{m}}^{\mathrm{2}}}$ ) 
& {32}\cite{Sundfors1974},  {9}\cite{Griffiths2019}   &  32\cite{Sundfors1974},  {9}\cite{Griffiths2019}   &  { 68}\cite{Sundfors1974},  {48}\cite{Griffiths2019}   & {2.1}\cite{Sundfors1974},  {5.3}\cite{Griffiths2019},1.9 \cite{Litvyak2021}   \\ 
\hline 
$S_{44}$ ($^{75}$As) / $S_{44}$($^{ }$ Ga) {\bf[this work] }  &    &     &    & {\bf 2.5$\pm$0.4}  \\ 
\hline
\hline 
\end{tabular}
\caption{Quadrupole parameters obtained in this work and those taken from the literature, as well as other GaAs parameters used in the calculations: stiffness tensor elements, isotope abundancies, quadrupole moments and gyromagnetic ratios. }
\label{tab:table1}
\end{table*}

It was assumed for many years that  the heat storage in bulk semiconductor NSS is dominated by dipole-dipole interaction.  Our recent measurements of the Overhauser field in the presence of the slowly varying external magnetic field overturned this assumption and pointed out possible contribution of strain-induced quadrupole interaction, see Eq.~(\ref{eq:BH}) \cite{Vladimirova2018}.  
The effective field due to quadrupole interaction can be calculated as  \cite{CH11Spin2017}:
\begin{equation} 
\label{eq:local} 
B_{Q}=\frac{4}{5} \frac{\sum_i {A_i(E_{Q\perp}^{i^2}+E_{QZ}^{i^2})}}
{\sum_i {A_i \gamma_{Ni}^{2}}}.
 \end{equation}
This yields $B_{Q}^{NMR}=16$~G using the quadrupole energies extracted from the NMR spectra. This value needs to be compared with $B_H$ from Eqs.~(\ref{eq:BH}, \ref{eqSpinT2}) that fits the re-magnetization experiments.

Figure~\ref{fig:demag} shows two sets of the spin noise spectra measured in a re-magnetization experiment starting from $B_i=0$ for two different initial nuclear spin temperatures obtained by optical pumping during $4$~minutes~(a) and $30$~s~(b).
The peak in the electron spin noise spectrum shifts from zero at $B=0$ to $\approx230$~MHz ($\approx100$~MHz) at $B_\perp=100$~G in the NSS cooled during $120$ ($30$) seconds. Longer  optical cooling yields lower NSS temperatures and higher Overhauser field reached in the former case, while the contribution from the external field is identical in these two experiments. It is shown by the yellow line.
One can see that the rapid growth of the frequency at small fields is followed by much slower growth at higher fields. The latter is essentially due to the external field. In the framework of the spin temperature theory, the characteristic magnetic field where the slope changes is of the order of heat capacity field $B_H$.
Red lines calculated using Eqs.~((\ref{eqSpinT0}), \ref{eq:BH}), (\ref{eq:freq}) and  (\ref{eqSpinT2})  assuming  $B_H=14\pm2$~G and $\Theta_{Ni}$ as a fitting parameter are shown in Fig.~\ref{fig:demag} on top of the color-encoded spin-noise spectra. The magnetic field dependent NSS relaxation on the time scale of the measurement ($50$~s) is taken into account in this calculation. The corresponding values of $1/T_1$ measured in a separate set of experiments are shown in the inset.
One can see that within the experimental accuracy, which is mainly given by the unavoidable warm-up of the NSS during measurements, $B_H\approx B_Q=14\pm2$~G matches quite well the value $B_Q^{NMR}=16$~G.  

Importantly, NMR and re-magnetization measurements were done at very close points on the sample surface. Indeed, it has been shown that the strain varies in the plane of GaAs substrates, epilayers, quantum wells and microstructures   \cite{Litvyak2021,Yamada1998,Eickhoff2003,Brunetti2007,Abbarchi2012,LastrasMartinez2017}. 
A different set of NMR experiments performed at another point yielded a smaller value of $B_Q^{NMR}=12$~G, correlated with smaller  $B_Q=9\pm2$~G extracted from the corresponding  re-magnetization experiments.

Note that the NSS relaxation enhancement at small fields has also been attributed to the presence of quadrupole effects in the similar samples \cite{Kotur2016,Vladimirova2017}. 
This so-called quadrupole relaxation is  induced by fluctuating donor charges, and the resulting enhancement of the relaxation rate as compared to the relaxation rate at strong field  $1/T_\infty$ is given by the same heat capacity field $B_H$, see Eq.~(\ref{eq:BH}): 
\begin{equation} 
\frac{1}{T_1}=\frac{1}{T_\infty}+\frac{1}{T_Q};  \qquad \frac{1}{T_Q}\propto1/(B^2+B_H^2).
\label{eq:TQ} 
 \end{equation}
Fitting the experimental values of the  NSS relaxation rate (Fig.~\ref{fig:demag}, inset) to the Eq.~(\ref{eq:TQ}) returns the value  $B_Q^r=9$~G which is slightly smaller, but still close to the the values returned by NMR and re-magnetization experiments.  
These results strongly support the ideas put forward in Refs.~\onlinecite{Kotur2016,Vladimirova2017,Vladimirova2018}.  Namely, the electric field gradients in the NSS containing isotopes with the spin $I>1/2$ lead to quadrupole effects which are detrimental to both nuclear spin memory and adiabatic demagnetization efficiency. 

\section{Estimation of gradient-elastic tensor elements}
In this section we further analyze the quadrupole parameters determined from NMR experiments and extract relevant information on the  stress intensity and orientation, as well as on the elements of the gradient-elastic tensor $S_{ijkl}$ - an important material parameter relating electric field gradients on the atomic sites with the strain experienced by the crystal.
In the crystals with cubic symmetry like GaAs, it can be reduced to two components, $S_{11}$ and $S_{44}$.
The interest in  this tensor  has emerged recently, both in view of non-destructive characterization of the strain, and in the context of exploration of coherent electron-nuclei spin dynamics in GaAs quantum dots.

The three isotope-dependent parameters that we extract from the NMR spectra are related to the three isotope-independent parameters which characterize the in-plane biaxial stress experienced by the crystal: values of the pressure applied in two orthogonal directions ($p_1$, $p_2$)  and the angle $\varsigma$ between the stress principal axis and  $[100]$ crystallographic axis. Their derivation is given in Section~\ref{sec:appendixQ} \footnote{The same formula wher used in Ref.~\onlinecite{Litvyak2021} }:
\begin{equation} 
\label{eq:EQZ} 
E_{QZ}^i=-\frac{3eQ^iS_{11}^i}{4I\left(2I-1\right)}\frac{p_1+p_2}{C_{11}-C_{12}},
 \end{equation}
 \begin{equation} 
 \label{eq:EQR} 
 E_{Q\perp}^i \mathrm{cos} 2\zeta^i=\frac{3\sqrt{3}eQ^iS_{11}^i}{4I\left(2I-1\right)}\frac{p_1-p_2}{C_{11}-C_{12}}{\mathrm{cos} \left(2\varsigma \right)\ },
\end{equation} 
\begin{equation} 
\label{eq:EQI} 
E_{Q\perp}^i\mathrm{sin} 2\zeta^i=\frac{\sqrt{3}eQ^iS_{44}^i}{2I\left(2I-1\right)}\frac{p_1-p_2}{C_{44}}{\mathrm{sin} \left(2\varsigma \right)\ }.
 \end{equation}
Here $e$ is the absolute value of the electron charge, $C_{11}$, $C_{12}$ and $C_{44}$ are known stiffness tensor components,  $E_{Qz}^i$, $E_{Q\perp}^i$ and $\zeta^i$ are the parameters of the Hamiltonian (\ref{eq:HQi}) determined from the fit to the NMR spectra. Their values are given in Table~\ref{tab:table1}.

Eqs. (\ref{eq:EQZ})-(\ref{eq:EQI}) are not sufficient to fully characterize five independent parameters ($p_1$, $p_2$, $\varsigma$, $S_{11}^i$, $S_{44}^i$  ) for each isotope.
However, they allow us to ascertain  the ratio between the values of gradient elastic tensor components  $S_{11}^{^{69}\mathrm{Ga}}/S_{11}^{^{75}\mathrm{As}}$ and $S_{44}^{^{69}\mathrm{Ga}}/S_{44}^{^{75}\mathrm{As}}$. 
Since both Ga isotopes experience the same electrostatic environment, these ratios are expected to be identical for $^{71}$Ga. This appears to be the case within our experimental precision.  In the following we denote both isotopes as Ga, to simplify the notations.

The obtained values of the gradient elastic tensor elements are summarized in Table~\ref{tab:table1}, and compared with previous results. 
The component $S_{11}$ of the gradient-elastic tensor contributes to both in and out-of-plane quadrupole energies, see Eqs~\ref{eq:EQZ}-\ref{eq:EQR}. In particular, it is proportional to $E_{QZ}$, that dominates over in-plane quadrupole splitting $E_{Q\perp}$ in our sample. One can see that  $S_{11}^{^{ }\mathrm{Ga}}/S_{11}^{^{75}\mathrm{As}}=-1.1\pm0.1$  matches quite well the conclusions of Refs. \onlinecite{Sundfors1974,Chekhovich2018,Litvyak2021}, corroborating their results. 

However for $S_{44}^{^{ }\mathrm{Ga}}/S_{44}^{^{75}\mathrm{As}}$ the discrepancy between the existing values is very substantial. Our  value $2.5\pm 0.4$ is rather close to  that of  Refs.~\onlinecite{Sundfors1974,Litvyak2021}, and much smaller that the result of Ref.~\onlinecite{Griffiths2019}.
%
At this stage we don't have any explanation for these discrepancies, but such a strong dispersion of $S_{44}$ measured in different experiments suggests that further studies are mandatory to resolve this issue. 

\section{Evaluation of the built-in  stress and disorder}
In order to estimate the stress parameters  $p_1$,  $p_2$ and $\varsigma$ Eqs.~(\ref{eq:EQZ})-(\ref{eq:EQI}) are not sufficient and we need to use the values of $S_{11}$ measured elsewhere. We rely on recent values of  Chekhovich et al \cite{Chekhovich2018}, \footnote{Using the values of Sundfors \cite{Sundfors1974} would give approximately a factor of  $1.4$ smaller values of   $p_1$ and $p_2$. }. 
For the point where the complete set of NMR spectra has been measured (see Fig.~\ref{fig:ColorMap}) this yields $p_1+p_2=29.5$~MPa, $p_1-p_2=14$~MPa, $\varsigma=55^\mathrm{o}$ degrees, bearing in mind that relative sign of $p_1+p_2$, $p_1-p_2$ and  $\varsigma$ is not uniquely defined. 
Measurements performed at other points of the sample revealed variation of the quadrupole energies up to $30$\%. 
This is consistent with the disorder effects reported in GaAs microcavities \cite{Brunetti2007,Abbarchi2012,LastrasMartinez2017}, as well as in bulk GaAs samples \cite{Yamada1998,Litvyak2021}. 
%
{
More specifically, in our microcavity sample  due to the difference of lattice constants between GaAs and AlAs plastic relaxation is likely to occur in the Bragg mirrors having $\approx 4$~$\mu$m thickness. This relaxation is eventually accompanied by the formation of dislocations. Therefore, the lattice constant of the bottom Bragg mirror would be different from that of GaAs cavity that we probe optically. It is not possible to give a quantitative estimation, but this difference should result in a strain in the cavity layer.
}



\section{Conclusions}
In this work we develop and implement a new technique that allows us to probe NMR  by electron spin noise spectroscopy.
These experiments  aim to quantify quadrupole effects in the NSS of n-GaAs. 
%
%
The NSS absorption spectra in a n-GaAs epilayer embedded in a microcavity are measured under magnetic field in the range from zero to $100$~G, either parallel or perpendicular to the growth axis. 
The ensemble of the spectra determines unambiguously the parameters of the quadrupole  Hamiltonian, and thus relevant  splittings between nuclear spin states for each of three GaAs isotopes.
These measurements are particularly relevant at zero and weak magnetic fields, because the spin states splittings  are dominated by quadrupole interaction rather than by Zeeman effect, and only few experimental methods give access to zero- and low-field NMR \cite{Blanchard2021,Litvyak2021}.

Our results establish a connection between  the residual strain giving rise to quadrupole splittings, the increased heat capacity and nuclear spin relaxation rates at low and zero field. 
Indeed, from the quadrupole parameters measured by NMR we estimate that heat capacity field  limiting NSS cooling by adiabatic demagnetization and increasing NSS spin relaxation rate is of order of $B_Q^{NMR}=16\pm2$~G.
This value is close to $B_Q=14\pm2$~G  measured at very close point on the sample surface in a set of separate adiabatic demagnetization experiments  using SNS as a detection tool.
More precise measurements are complicated  due to  variation up to 30\% of the  quadrupole  energies and thus nuclear warm-up rates induced by inhomogeneity of the local strain on the scale of several millimeters across the sample surface.
Nevertheless our results strongly support the model of quadrupole-limited  NSS cooling \cite{Vladimirova2018} and quadrupole-driven NSS warm-up \cite{Kotur2016,Vladimirova2017}. 

From the quadrupole hamiltonian we also obtain an estimation of the gradient-elastic tensor that relates gradients of the electrostatic potential on each nuclear site with the strain tensor.
Since we don't have an independent measurement of the stress present in the sample, we are limited to the determination of only relative values of the two relevant gradient-elastic tensor elements for different isotopes, $S_{11}$ and  $S_{44}$.
It appears that  $S_{11}^{^{ }\mathrm{Ga}}/S_{11}^{^{75}\mathrm{As}}=-1.1\pm0.1$  matches quite well the conclusions of Refs. \onlinecite{Sundfors1974,Chekhovich2018,Litvyak2021}, while for $S_{44}^{^{ }\mathrm{Ga}}/S_{44}^{^{75}\mathrm{As}}$, where a big discrepancy exist between the  values in the literature, we get  $S_{44}^{^{ }\mathrm{Ga}}/S_{44}^{^{75}\mathrm{As}}=2.5\pm 0.4$, rather close to  that of  Refs.~\onlinecite{Sundfors1974,Litvyak2021}, and much smaller than the result of Ref.~\onlinecite{Griffiths2019}. 
Understanding of  discrepancies in the $S_{44}$ values involved in the presence of an in-plane shear strain calls for further studies.

%

\section{Acknowledgements}

The authors are grateful to R.~Cherbunin and V.~K.~Kalevich for enlightening discussions, and acknowledge financial support from Russian Foundation for Basic Research grant N 22-42-09020, French national research agency grant ANR-21-CE30-0049,  French Embassy in Moscow (Ostrogradski fellowship for young researchers 2020), and French RENATECH network.

\section{Appendix }
\subsection{Determination of the NSS absorption from the electron spin noise peak frequencies. }
\label{sec:appendixA}

An OMF $B_1 \mathrm{cos}( 2\pi ft)$ along $z$-axis creates an energy flux towards NSS,  $Q(f, t)$, corresponding to the energy change per one spin given by
\begin{equation}
Q(f,t)=\frac{\partial E_{NSS}}{\partial t} =\frac{dB_1}{d t} \sum_{i}{A_i h \gamma_{Ni} I_{zi}},
\label{eq:qw1}
\end{equation}
where index $i$ runs over all isotopes and $I_{zi}$ is an average  projection of the $i$-th isotope spin on the OMF direction
\begin{equation}
I_{zi}=\frac{B_1}{h\gamma_{Ni}}  \left[
\chi_f'\mathrm{cos}(2\pi f t)-\chi_f''\mathrm{sin}(2\pi f t)
\right].
\label{eq:Iz}
\end{equation}
Here  $\chi_f'=\chi_{-f}'$  and $\chi_f''=-\chi_{-f}''$  are real  and imaginary  parts of the NSS susceptibility, respectively.
Therefore, the averaged over a period energy flux towards NSS $Q(f)$ is related to its absorption    $\mathcal{A}(f)=f\sum_{i}{A_i\chi_{fi}''}$  as 
\begin{equation}
Q(f)=\pi  B_1^2\mathcal{A}(f).
\label{eq:qw2}
\end{equation}

On the other hand the warm-up rate of the NSS in the presence of the OMF  reads \cite{OO}:
\begin{equation}
\frac{1}{T_w(f)}=\frac{ k_B\Theta_N}{C_N}Q(f)=\pi B_1^2\frac{ k_B\Theta_N}{C_N}\mathcal{A}(f).
\label{eq:Tw}
\end{equation} 
Here $C_N$ is the heat capacity of the NSS
\begin{equation}
C_N=\frac{I(I+1)}{3 }\left( B^2+B_H^2\right)h^2 \sum_{i}{A_i h \gamma^2_{Ni}}, 
\label{eq:CN}
\end{equation}
$B$ is an arbitrary external static field and  $B_H$ is the heat capacity field introduced in Eq.~(\ref{eq:BH}). It can be  rigorously determined as 
\begin{equation}
B_H^2=\frac{3 }{I(I+1)}\frac{Sp\left (\hat{H}^2_{SS} \right )}{\sum_{i}{A_i h \gamma^2_{Ni}}}, 
\label{eq:BH2}
\end{equation}
where $\hat{H}_{SS}$ is the total Hamiltonian of all nuclear interactions excluding Zeeman \cite{CH11Spin2017}.
%
%
Thus, NSS absorption at a given spin temperature is proportional to the rate of NSS  warm-up induced by an OMF.     

$T_w(f)$ can be extracted from the experiments described in Section~\ref{sec:protocol} in a following way.
Following Ref.~\cite{OO}, Ch.~5 we can write a detailed balance equation for the inverse nuclear spin temperature $\beta=1/k_B\Theta_N$ in the presence of the OMF:
\begin{equation}
{\dot{\beta}} =- \frac{1}{T_t(f)}\left( \beta-\beta_t \right ),
\label{eq:beta1}
\end{equation}
where 
\begin{equation}
\frac{1}{T_t(f)}=\frac{1}{T_w(f)}+\frac{1}{T_1}
\label{eq:tt}
\end{equation}
is the total relaxation rate including  OMF-independent relaxation at rate $T_1$, $\beta_t=\beta_0 \left ( 1+ T_w(f)/T_1 \right )^{-1}$ is the steady-state spin temperature and $\beta_0$ is the steady-state spin temperature for $B_1=0$.

Nuclear spin temperature is related to the Overhauser field-induced spin noise frequency component $F_N$ via Eqs.~(\ref{eqSpinT0}) and (\ref{eq:freq}). Therefore, from Eqs.~(\ref{eq:beta1}) we obtain 
\begin{equation}
\frac{1}{T_t(f)}=\frac{1}{\Delta t_6}\mathrm{log} \left( \frac{F_N(t_7)}{F_N(t_5)} \right),
\label{eq:tt1}
\end{equation}
where $\Delta t_6$ is the duration of the $6$-th stage of the NMR experiment, and $F_N(t_5)$ and $F_N(t_7)$ are frequencies measured at $5$-th and $7$-th stages of the NMR experiment, respectively.

Similarly, if  $B_1=0$ Eq.~(\ref{eq:beta1}) becomes
\begin{equation}
{\dot{\beta}} =- \frac{1}{T_1}\left( \beta-\beta_0 \right ),
\label{eq:beta0}
\end{equation}
and 
\begin{equation}
\frac{1}{T_1}=\frac{1}{\Delta t_6}\mathrm{log} \left( \frac{F^0_N(t_7)}{F^0_N(t_5)} \right ),
\label{eq:t11}
\end{equation}
where $F_N^0(t_5)$ and $F_N^0(t_7)$ are frequencies measured at $5$-th and $7$-th stages of the identical experiment but without OMF ($B_1=0$).

$1/T_w(f)$ is recalculated from  Eqs.~(\ref{eq:tt}), (\ref{eq:tt1}) and (\ref{eq:t11}) for each OMF frequency.
Since in high-temperature approximation  $\mathcal{A}(f) \propto \beta$, $1/T_w(f)$ does not depend on temperature and thus pertinently characterizes NSS absorption.
Therefore, $1/\pi B_1^2 T_w(f)$ constitutes an NMR spectrum, or, equivalently, NSS absorption spectrum. 
Such spectra, normalized to unity and encoded in colors, are shown in Fig.~\ref{fig:ColorMap}~(a, b) for various values of static magnetic field.

\subsection{Relation between the in-plane stress and  energetic parameters of the quadrupole Hamiltonian. }
\label{sec:appendixQ}
%
The second rank tensors describing stress $\sigma _{ij}$ and strain $\varepsilon _{kl}$ are related by the forth rank elasticity tensor ${C_{ijkl}}$:
\begin{equation}
{\sigma _{ij}} = \sum_{k,l}{C_{ijkl}}{\varepsilon _{kl}}.
\label{eq:elasticity}
\end{equation}
In cubic crystals there are only three nonzero components of the elasticity tensor:
\begin{align}
&{C_{11}} = {C_{xxxx}} = {C_{yyyy}} = {C_{zzzz}}; \nonumber  \\
&{C_{12}} = {C_{xxyy}} = {C_{yyxx}} = {C_{xxzz}} = {C_{zzxx}} = {C_{yyzz}} = {C_{zzyy}}; \nonumber \\
&{C_{44}} = {C_{xyxy}} = {C_{yzyz}} = {C_{xzxz}}.  
\label{eq:C}
\end{align}
In a crystalline plate with the normal ($z$-axis) parallel to the $[001]$ axis, application of the pressure $p$ in the structure plane at the angle $\varsigma $ to the $[100]$ axis results in appearance of the following components of the stress tensor: 
\begin{align}
&{\sigma _{xx}} = p{\cos ^2}\varsigma \nonumber   \\
&  {\sigma _{yy}} = p{\sin ^2}\varsigma  \nonumber  \\
&  {\sigma _{xy}} = {\sigma _{yx}} = p\cos \varsigma \sin \varsigma \nonumber  \\
&  {\sigma _{zz}} = {\sigma _{zx}} = {\sigma _{xz}} = {\sigma _{zy}} = {\sigma _{yz}} = 0  
\label{eq:sigma}
\end{align}
By expressing components of the stress tensor in Eq.~(\ref{eq:sigma}) via strain components with Eqs.~(\ref{eq:elasticity}) and (\ref{eq:C}), one obtains a system of linear equations for strain tensor components that yields the following solutions (see also problem $3$ to $10$th section of Chapter 1 in \cite{LL_V7}): 
\begin{align}
  {\varepsilon _{xx}} &= p \cdot \frac{{\left( {{C_{11}} + 2{C_{12}}} \right){{\cos }^2}\varsigma  - {C_{12}}}}{{\left( {{C_{11}} - {C_{12}}} \right)\left( {{C_{11}} + 2{C_{12}}} \right)}}  \nonumber  \\
  {\varepsilon _{yy}} &= p \cdot \frac{{\left( {{C_{11}} + 2{C_{12}}} \right){{\sin }^2}\varsigma  - {C_{12}}}}{{\left( {{C_{11}} - {C_{12}}} \right)\left( {{C_{11}} + 2{C_{12}}} \right)}}  \nonumber  \\
  {\varepsilon _{zz}} &=  - p \cdot \frac{{{C_{12}}}}{{\left( {{C_{11}} - {C_{12}}} \right)\left( {{C_{11}} + 2{C_{12}}} \right)}}  \nonumber \\
  {\varepsilon _{zx}} &= {\varepsilon _{xz}} = {\varepsilon _{zy}} = {\varepsilon _{yz}} = 0  \nonumber  \\
  {\varepsilon _{xy}} &= {\varepsilon _{yx}} = p \cdot \frac{{\cos \varsigma \sin \varsigma }}{{{C_{44}}}}. 
\label{eq:epsilon}
\end{align}
In the general case, the stress can be applied along two orthogonal axes in the plane. If the angle $\varsigma $ defines the direction of one of the two principal axes of the stress tensor in the plane, principal values of this tensor being ${p_1}$ and ${p_2}$, then due to linearity of  Eqs.~(\ref{eq:epsilon}) we get: 
\begin{align}
    {\varepsilon _{xx}} &= \frac{{\left( {{C_{11}} + 2{C_{12}}} \right)\left( {{p_1}{{\cos }^2}\varsigma  + 
    {p_2}{{\sin }^2}\varsigma } \right) - {C_{12}}\left( {{p_1} + {p_2}} \right)}}{{\left( {{C_{11}} - {C_{12}}} \right)\left( {{C_{11}} + 2{C_{12}}} \right)}}   \nonumber  \\
  {\varepsilon _{yy}} &= \frac{{\left( {{C_{11}} + 2{C_{12}}} \right)\left( {{p_1}{{\sin }^2}\varsigma  
  + {p_2}{{\cos }^2}\varsigma } \right) - {C_{12}}\left( {{p_1} + {p_2}} \right)}}{{\left( {{C_{11}} - {C_{12}}} \right)\left( {{C_{11}} + 2{C_{12}}} \right)}}    \nonumber \\
  {\varepsilon _{zz}} &=  - \frac{{{C_{12}}\left( {{p_1} + {p_2}} \right)}}{{\left( {{C_{11}} - {C_{12}}} \right)\left( {{C_{11}} + 2{C_{12}}} \right)}}  \nonumber  \\
  {\varepsilon _{zx}} &= {\varepsilon _{xz}} = {\varepsilon _{zy}} = {\varepsilon _{yz}} = 0   \nonumber   \\
  {\varepsilon _{xy}} &= {\varepsilon _{yx}} = \left( {{p_1} - {p_2}} \right) \cdot \frac{{\cos \varsigma \sin \varsigma }}{{{C_{44}}}}. 
\label{eq:epsilon2}
\end{align}
Thus, with the knowledge of the strain components one can find the energetic parameters of the quadrupole Hamiltonian  \cite{Abragam}: 
\begin{align}
{E_{QZ}} &= \frac{{3eQ{S_{11}}}}{{2I\left( {2I - 1} \right)}}\left( {{\varepsilon _{zz}} - \frac{{{\varepsilon _{xx}} + 
{\varepsilon _{yy}} }}{3}} \right)    \nonumber  \\
  {E_{QR}} &= \frac{{3\sqrt 3 eQ{S_{11}}}}{{4I\left( {2I - 1} \right)}}\left( {{\varepsilon _{xx}} - {\varepsilon _{yy}}} \right)    \nonumber  \\
  {E_{QI}} &= \frac{{\sqrt 3 eQ{S_{44}}}}{{2I\left( {2I - 1} \right)}}\left( {{\varepsilon _{xy}} + {\varepsilon _{yx}}} \right)   
\label{eq:EQeps}
\end{align}
which amount to ({\it cf} Eqs.~(\ref{eq:HQi}) and (\ref{eq:EQZ})-(\ref{eq:EQI})):
\begin{align}
{E_{QZ}} &=  \frac{{3eQ{S_{11}}}}{{4I\left( {2I - 1} \right)}}\frac{1}{{{C_{11}} - {C_{12}}}}\left( {{p_1} + {p_2}} \right)  \nonumber \\
{E_{QR}} &= \frac{{3\sqrt 3 eQ{S_{11}}}}{{4I\left( {2I - 1} \right)}}\frac{{\cos \left( {2\varsigma } \right)}}{{{C_{11}} - {C_{12}}}}\left( {{p_1} - {p_2}} \right) \nonumber  \\
  {E_{QI}} &= \frac{{\sqrt 3 eQ{S_{44}}}}{{2I\left( {2I - 1} \right)}}\frac{{\sin \left( {2\varsigma } \right)}}{{{C_{44}}}}\left( {{p_1} - {p_2}} \right)
\label{eq:EQp}
\end{align}
\bibliography{refsNSS}

\end{document}